\documentclass[12pt]{article}
\usepackage{amsmath}
\usepackage{amsthm}
\usepackage{amssymb}
\usepackage{mathrsfs}
\usepackage{authblk}
\usepackage[usenames]{color}
\usepackage[utf8]{inputenc}
\usepackage{cite}
\usepackage{pdfsync}
\usepackage{enumitem}
\setlist{parsep=0pt,itemindent=0pt}

\usepackage{graphicx}

\theoremstyle{definition}

\numberwithin{equation}{section}
\numberwithin{thm}{section}
\numberwithin{lemma}{section}
\numberwithin{prop}{section}
\numberwithin{cor}{section}
\numberwithin{rmk}{section}
\numberwithin{defn}{section}
\numberwithin{exa}{section}

\newcommand{\dx}{\partial_x}
\newcommand{\dy}{\partial_y}

\newcommand{\dt}{\partial_t}

\newcommand{\dw}{\partial_w}

\newcommand{\gen}[1]{\partial_{#1}}

\begin{document}
\pagenumbering{arabic}
\clearpage
\thispagestyle{empty}

\title{Comment on "An optimal system, invariant solutions, conservation laws, and complete
	classification of Lie group symmetries for a generalized (2+1)-dimensional
	Davey--Stewartson system of equations for the wave propagation in water of finite
	depth"}

\author[1]{F.~G\"ung\"or\thanks{gungorf@itu.edu.tr}}

\author[2]{C. Özemir \thanks{ozemir@itu.edu.tr}}

\affil[1]{Department of Mathematics, Faculty of Science and
Letters, Istanbul Technical University, 34469 Istanbul, Turkey}

\date{\today}
\maketitle

\abstract{
In this  comment we report that a recent paper [{\em Eur. Phys. J. Plus} 138, 195 (2023)] contains significant errors, omissions and inaccuracies.}

\section{Introduction}

In Ref. \cite{DhimanKumar2023}  the so called generalized Davey--Stewartson (DS) system in (2+1)-dimensions in the form
\begin{subequations}\label{IDS}
	\begin{eqnarray}
		\label{makdenka}  iZ_t+\frac{\alpha^2}{2}(Z_{xx}+\alpha^2 Z_{yy})-\nu|Z|^2Z+Z  Q_x&=&0, \\
		\label{makdenkb}  Q_{xx}-\alpha^2 Q_{yy}-2\nu(|Z|^2)_{x}&=&0
	\end{eqnarray}
\end{subequations}
where $\nu=\mp 1$ and $\alpha^2=\mp 1$, has been considered from the symmetry point of view. On the contrary to the authors' claim, neither the study of this system in this context is entirely authentic nor their results are novel. In fact, system \eqref{IDS} is the integrable potential variant of the standard  DS system with the $x, y$ coordinates interchanged \cite{ChampagneWinternitz1988}
\begin{subequations}\label{DSCW}
	\begin{eqnarray}
		\label{DSCWa}  i\psi_t+\psi_{xx}+\epsilon_1 \psi_{yy}&=&\epsilon_2 |\psi|^2\psi+\psi  w, \\
		\label{DSCWb}  w_{xx}+\delta_1 w_{yy}&=&\delta_2(|\psi|^2)_{yy},
	\end{eqnarray}
\end{subequations}
where  $\epsilon_1=\mp 1$, $\epsilon_2 = \mp 1$, and $\delta_1$, $\delta_2$ are constants. This is easily seen by choosing the parameters as
\begin{equation}\label{param}
 \epsilon_1=-\delta_1=\alpha^2, \quad \epsilon_2=\nu,  \quad \delta_2=2\nu\alpha^2,
\end{equation}
exchanging the spatial coordinates in \eqref{DSCW} with the scaling   $(x,y)\to  1/\sqrt{2}(y,x)$ and setting $\psi=Z$, $w=-Q_x$.
Lie symmetry algebra, symmetry group, one and two-dimensional subalgebras and symmetry reductions together with solutions of the Davey--Stewartson system  \eqref{DSCW} was  investigated in the late eighties in \cite{ChampagneWinternitz1988}.   Lie symmetry algebra of \eqref{DSCW} was shown to be an infinite-dimensional  Kac--Moody--Virasoro algebra  in the integrable case $\delta_1=-\epsilon_1$. In this case, the symmetry algebra includes 4 arbitrary functions of time. If $\delta_1\neq -\epsilon_1$, the Lie algebra is still an infinite-dimensional, which is a Kac--Moody algebra including 3 arbitrary functions of time. After this work several extensions   have been made like a generalized DS system \cite{GuengoerAykanat2006, LiYeChen2008} in a different medium, variable-coefficient DS system \cite{GuengoerOezemir2016}, (3+1)-dimensional DS system \cite{Oezemir2020} within the same framework. Its dispersionless case can be found in \cite{GuengoerOezemir2021}.

Lie symmetry algebra of  system \eqref{IDS} depending on all four parameters (inclusive the integrable version) has already been found  by one of the present authors  in a joint work \cite{GoenuelOezemir2022}, where the  system  considered was
\begin{subequations}\label{DSGO}
	\begin{eqnarray}
		\label{DSCWa}  i\psi_t+\psi_{xx}+\epsilon_1 \psi_{yy}&=&\epsilon_2 |\psi|^2\psi+\psi  w_y, \\
		\label{DSCWb}  w_{xx}+\delta_1 w_{yy}&=&\delta_2(|\psi|^2)_{y},
	\end{eqnarray}
\end{subequations}
where $\epsilon_1=\mp 1$, $\epsilon_2 = \mp 1$, and $\delta_1$, $\delta_2$ are constants.
Using the amplitude-phase representation of the wave in the form $\psi(x,y,t)=M(x,y,t)e^{i\sigma(x,y,t)}$,  its infinite-dimensional Lie symmetry algebra   is generated by the vector field
\begin{equation}
\mathcal{V}=\mathcal{X}(\tau)+\mathcal{Y}(\alpha)+\mathcal{Z}(\beta)+\mathcal{W}(\mu)+\mathcal{R}(\gamma)+\mathcal{S}(\delta)
\end{equation}
with
\begin{equation}\label{gen}
	\begin{aligned}
		\mathcal{X}(\tau)&=\tau(t)\dt+\frac{\dot \tau(t)}{2} (x\dx+y\dy-M\gen M-w\dw)+    \\
		&+\frac{x^2+ \epsilon_1 y^2}{8}\ddot{\tau}(t)\gen \sigma
		-\frac{\dddot\tau(t)}{8}(x^2y+\frac{\epsilon_1}{3} y^3)\dw,\\
		\mathcal{Y}(\alpha)&=\alpha(t)\dx+\frac{x}{2}\dot{\alpha}(t)\gen \sigma-\frac{xy}{2}\ddot{\alpha}(t)\dw,\\
		\mathcal{Z}(\beta)&=\beta(t)\dy+\frac{ y}{2\epsilon_1}\dot{\beta}(t)\gen \sigma+\frac{\ddot{\beta}(t)}{4 \epsilon_1}(\delta_1x^2- y^2)\dw,\\
		\mathcal{W}(\mu)  &=\mu(t) \gen \sigma-y\dot\mu(t)\dw, \\
		\mathcal{R}(\gamma)& = x \gamma(t) \dw,\\
		\mathcal{S}(\delta)&= \delta(t)\dw.
	\end{aligned}
\end{equation}
where $\alpha(t)$, $\beta(t)$, $\mu(t)$, $\gamma(t)$, $\delta(t)$ are arbitrary functions and the constants $\epsilon_1$, $\delta_1$ and the function $\tau(t)$ are  to satisfy
\begin{equation}
	(\epsilon_1+\delta_1)\dddot{\tau}(t)=0,
\end{equation}
giving the result
\begin{equation}
	\tau(t) = \begin{cases}
		\text{arbitrary,} &  \delta_1=-\epsilon_1=\mp1, \\
		\tau_2 t^2+\tau_1 t+\tau_0, &  \delta_1 \neq -\epsilon_1.
	\end{cases}
\end{equation}		
In order to obtain the symmetry algebra of \eqref{IDS} it is sufficient to make a simple change of scaling and to exchange the roles of $x$ and $y$  in the symmetry algebra obtained in \cite{GoenuelOezemir2022}.

The main deficiency of \cite{DhimanKumar2023} is that the authors  treat the subalgebra restricted by setting three of the six arbitrary functions to  linear functions and  to a constant  and attempt to obtain  one-dimensional  subalgebras as if it is a finite-dimensional algebra.  Classification of one- and two-dimensional subalgebras of the DS symmetry algebra can be found in \cite{ChampagneWinternitz1988}. Methods exist in the literature for classification of subalgebras of  infinite-dimensional Lie algebras. For example, we refer to  a recent work\cite{MaltsevaPopovych2021} for details. Symmetry reduced (1+1)-dimensional system under the erroneously obtained one-dimensional subalgebras are then subjected to a further symmetry procedure to reduce them to a system of ODEs. The DS system is known to possess soliton and multi-solution solutions. On the other hand, the solutions claimed to be new in \cite{DhimanKumar2023} are far from satisfying the system \eqref{IDS}. First of all, the real-valued amplitude $u=|Z|$ of the wave $Z$ can not vanish.  However, $u$ components of the solutions in formulas  (49), (58) and (67) of \cite{DhimanKumar2023} are zero.  The reality of  the amplitudes $u$   in formulas (48), (57) and (66) of \cite{DhimanKumar2023} is not satisfied if $\nu=1$. In the case $\nu=-1$, the solution presented in formula (57)   does not satisfy \eqref{IDS}. Furthermore, with the aid of Mathematica symbolic package we checked the remaining presented solutions (26), (33), (40) of \eqref{IDS}    and noticed that none of them satisfy this system.

Finally, we mention that a well-known fact concerning conservation laws of the integrable DS system is that they admit an infinite hierarchy of conservation laws (See for example  \cite{Omote1988}).

\bibliographystyle{unsrt}	

\end{document}